\documentclass[12pt]{article}
\usepackage{graphicx}
\usepackage{amssymb}
\usepackage{amsmath}
\usepackage{color}

\newcommand{\bear}{\begin{array}}  
\newcommand {\eear}{\end{array}}
\newcommand{\bea}{\begin{eqnarray}}   
\newcommand{\eea}{\end{eqnarray}}
\newcommand{\beq}{\begin{equation}}   
\newcommand{\eeq}{\end{equation}}
\newcommand{\bef}{\begin{figure}}  \newcommand 
{\eef}{\end{figure}}
\newcommand{\bec}{\begin{center}}  \newcommand 
{\eec}{\end{center}}

\begin{document}

\begin{titlepage}

\begin{flushright}
ICRR-Report-536 \\
IPMU 09-0008 \\
TU-837 \\
June, 2009
\end{flushright}

\vskip 1cm

\begin{center}
{\large \bf
Cosmic Rays from Dark Matter Annihilation and \\Big-Bang Nucleosynthesis
}
\vskip 1cm

Junji Hisano$^{(a,b)}$,
Masahiro Kawasaki$^{(a,b)}$,
Kazunori Kohri$^{(c)}$,
Takeo Moroi$^{(b,d)}$ and
Kazunori Nakayama$^{(a)}$
 
\vskip 0.5cm

{\it $^a$Institute for Cosmic Ray Research,
University of Tokyo, Kashiwa 277-8582, Japan}

\vskip 0.3cm

{\it $^b$Institute for the Physics and Mathematics of the Universe,
University of Tokyo, Kashiwa 277-8568, Japan}

\vskip 0.3cm

{\it $^c$Physics Department, Lancaster University, Lancaster LA1 4YB, UK}

\vskip 0.3cm

{\it $^b$Department of Physics, Tohoku University, Sendai 980-8578, Japan}

\begin{abstract}
  Recent measurements of cosmic-ray electron and positron fluxes by
  PAMELA and ATIC experiments may indicate the existence of
  annihilating dark matter with large annihilation cross section.  We
  show that the dark matter annihilation in the big-bang
  nucleosynthesis epoch affects the light element abundances, and it
  gives stringent constraints on such annihilating dark matter
  scenarios for the case of hadronic annihilation. 
  Constraints on leptonically annihilating dark matter models are less severer.
\end{abstract}

\end{center}
\end{titlepage}

\section{Introduction}

Cosmological observations have revealed that about 20 percent of the
total energy density of the Universe is dominated by the dark matter
(DM) \cite{Komatsu:2008hk}, whose detailed properties are still
unknown, and many physicists believe that the DM is a kind of stable
particle appearing in the physics beyond the standard model.  Thus,
to determine the origin and nature of the dark matter in the 
Universe is one of the most important topics in the particle physics,
and some methods were proposed for detecting the signals of the DM
directly or indirectly \cite{Jungman:1995df,Bertone:2004pz}.  One of
such methods is that to search for high-energy cosmic-rays, including
gamma-rays, positrons, anti-protons and neutrinos, which come from the
DM annihilation in our Galaxy.

Recent results of the cosmic-ray positron and electron fluxes by the
PAMELA satellite experiment \cite{Adriani:2008zr} and the ATIC balloon
experiment \cite{:2008zz} are now drawing a lot of attention, since
the steep excess observed by these experiments can be interpreted as
an extra contribution from the DM annihilation.  (For earlier papers,
see \cite{DMannihilation}.)  However, in order to explain these
signals the annihilation cross section should be fairly large, as
$\langle \sigma v\rangle \sim (10^{-24}-10^{-23})~{\rm cm^3s^{-1}}$,
which may be achieved by the Sommerfeld enhancement effect
\cite{Hisano:2003ec}.  This is orders of magnitude larger than the
standard value $\langle \sigma v\rangle \sim 3\times10^{-26}~{\rm cm^3
  s^{-1}}$, which reproduces the observed DM abundance under the
thermal freezeout scenario \cite{Jungman:1995df}. Then, a huge boost
factor $(B_F\sim 100)$, which is due to the enhancement of the DM
annihilation rate due to the clumpy structure of the DM halo, should
be introduced in the scenario if $\langle \sigma v\rangle$ is not
varied in time.  However, the DM may be produced nonthermally in other
ways, such as late-decay of long lived particles
\cite{Kawasaki:1995cy,Moroi:1999zb}.  Once we give up the thermal
freezeout scenario, a large annihilation cross section is still
allowed.

The DM with large annihilation rate leads to other observable
signatures: gamma-rays, anti-protons, and neutrinos.\footnote
{Indirect detection signatures of non-thermally produced DM were
  investigated in Ref.~\cite{Profumo:2004ty} before PAMELA and ATIC.}
In particular, even if the DM only annihilates into leptons, the
internal bremsstrahlung processes always emit significant amount of
gamma-rays, and it also predicts a comparable amount of neutrinos.
Those may put stringent constraints on the DM models
\cite{Bertone:2008xr, Hisano:2008ah}.

Besides these cosmic-ray signals, in this paper we show that the DM
annihilation also affects the prediction of the big-bang
nucleosynthesis (BBN), since it injects high-energy particles in the
nucleosynthesis epoch, which modify the light elements abundances.
Such an effect of DM annihilation was pointed out by Jedamzik in
Ref.~\cite{Jedamzik:2004ip}, where the modification on the $^6$Li
abundance due to the hadronic process was emphasized.  (See also
Refs.~\cite{Reno:1987qw,Frieman:1989fx} for early studies of DM
annihilation effects on BBN.)  This subject was recently studied by
four of the present authors in connection with the observed positron
excess \cite{Hisano:2008ti}.  In this paper we have performed more
systematic studies on the effect of DM annihilation on BBN, including
the case where the DM annihilates into only leptons, motivated by
recent results of PAMELA/ATIC.  
In such a case, the
photo-dissociations of light elements give constraint.  In
particular, we found that the $^3$He to D ratio gives the most
stringent constraint on the annihilation cross section, which  can be
consistent with the PAMELA/ATIC results.  
We also consider
the case that the DM annihilates into hadrons.  Then, the BBN
constraint becomes more stringent, and a boost factor larger than
unity must be introduced in order to account for the PAMELA/ATIC
anomalies.  Therefore, the DM annihilation models as an explanation of the
PAMELA/ATIC results should be treated carefully so as not to
contradict with the BBN constraint presented in this paper.

This paper is organized as follows.  In Sec.~\ref{sec:AnnBBN} effects
of DM annihilation on BBN are explained both for the cases of
radiative and hadronic annihilation modes.  Current status of
observations of light element abundances is summarized there.  In
Sec.~\ref{sec:constraint} the resultant constraints on the DM
annihilation cross section and also their implications to the
PAMELA/ATIC observations are presented.  Sec.~\ref{sec:conc} is
devoted to conclusions and discussion.

\section{BBN with Annihilating Dark Matter}
\label{sec:AnnBBN}

\subsection{Light element abundances: theory}

As discussed in the introduction, we consider the BBN scenario with
DM particle which has sizable annihilation cross section.  In
such a scenario, the annihilation rate of the DM during the BBN epoch
may be significant and the abundances of the light elements may be
affected by energetic particles emitted via the annihilation process.

For the study of the effects of the DM annihilation during BBN epoch,
it is necessary to calculate the production rate of energetic
particles (i.e., $\gamma$, $e^\pm$, hadrons, and so on) via the
annihilation.  Denoting the pair annihilation cross section of the DM
as $\langle\sigma v\rangle$, the production rate of the energetic
particle $I$ is given by
\begin{eqnarray}
  \left[ \frac{d f_I}{d t} \right]_{\rm ann} = 
  \frac{1}{2}
  n_{\rm DM}^2 \langle \sigma v \rangle 
  \left[ \frac{d N_I}{d E} 
  \right]_{\rm ann},
  \label{fdot}
\end{eqnarray}
where $f_I(E,t)$ is the energy spectrum of the particle $I$ at a time
$t$, and $\left[dN_I/dE\right]_{\rm ann}$ is the energy distribution
of $I$ from the single annihilation process.  In our study, we assume
that the pair annihilation is dominantly via $s$-wave processes, and
that $\langle\sigma v\rangle$ is independent of time.

In addition, $n_{\rm DM}$ is the DM number density (at the time $t$),
which is given by
\begin{eqnarray}
  n_{\rm DM} = \frac{\Omega_{\rm DM} \rho_{\rm crit}}{m_{\rm DM}}
  \left( \frac{a_0}{a(t)} \right)^3,
\end{eqnarray}
where $\rho_{\rm crit}$ is the present critical density, $m_{\rm DM}$
is the DM particle mass, and $a_0$ and $a(t)$ are the scale factors at
present and at the time $t$, respectively.  For the density parameter
for the DM, we take $\Omega_{\rm DM}=0.206$ \cite{Komatsu:2008hk}.  The
energy distribution $\left[dN_I/dE\right]_{\rm ann}$ strongly depends
on the properties of the DM; it is sensitive to how the DM
annihilates.  In particular, if quarks and/or gluon are produced via
the annihilation, we should take account of the hadronization process
in calculating $\left[dN_I/dE\right]_{\rm ann}$.  We use the PYTHIA
package \cite{Sjostrand:2006za} for the precise estimation of the
distribution functions.

It should be noted that, as one can see from Eq.\ \eqref{fdot}, the
production rate of the energetic particles (per one DM particle),
which is given by $\sim n_{\rm DM}\langle \sigma v\rangle$, is
proportional to $a^{-3}$ and is enhanced in the early Universe.  This
is a significant contrast to the case where the DM particle is
unstable.  The PAMELA and ATIC anomalies can be explained if the
present production rate of $e^\pm$ is $\sim 10^{-26}\ {\rm s}^{-1}$,
which may be realized either by the annihilation or the decay.  If the
production of $e^\pm$ is via the decay, the production rate is given
by the decay rate of the DM particle and is a constant of time.  Then,
the production rate of the energetic particles is too small during the
BBN epoch to affect the abundance of light elements.

Effects of the DM annihilation is classified into (i)
photo-dissociation, and (ii) $p\leftrightarrow n$ conversion due to
emitted pions and hadro-dissociation.  In the following, we summarize
important points in these effects.

\subsubsection{Photo-dissociation of light elements}

The high energy charged leptons and photons emitted into the cosmic
plasma induce electromagnetic showers and produce copious energetic
photons.  These photons destroy the light elements ($^4$He, $^3$He, D
and $^7$Li) synthesized in BBN through photo-dissociate
processes~\cite{RadDec,Kawasaki:1994sc}.

The effect of photo-dissociation is determined only by the amount of
total visible energy $E_{\rm vis}$ of produced particles in the
annihilation if $E_{\rm vis} \gg 10$~MeV \cite{RadDec}. Here ``visible
energy'' is defined by the sum of energies carried by $e^\pm$s and
photons after annihilation and subsequent decay.  Importantly, $E_{\rm
  vis}$ depends on the annihilation modes of dark matter.  We estimate
$E_{\rm vis}$ using the PYTHIA package; for the final states which
will be studied in the following, we show the ratio $E_{\rm
  vis}/m_{\rm DM}$ in Table \ref{table:evis}.  (Notice that the ratio
$E_{\rm vis}/m_{\rm DM}$ is independent of $m_{\rm DM}$ once the final
state is fixed.  In addition, $E_{\rm vis}\leq 2m_{\rm DM}$ because we
consider pair annihilation processes of the DM.)

\begin{table}
  \begin{center}
    \begin{tabular}{ll}
      \hline \hline
      Final State & $E_{\rm vis}/m_{\rm DM}$ 
      \\
      \hline
      $e^+ e^-$        & 2.00 \\
      $\mu^+ \mu^- $   & 0.70 \\
      $\tau^+ \tau^- $ & 0.62 \\
      $W^+ W^-$        & 0.94 \\      
      \hline \hline
    \end{tabular}
    \caption{\small Visible energy carried by $e^\pm$s and photons
      after annihilation and subsequent decay, $E_{\rm vis}$, in
      single DM annihilation process.}
    \label{table:evis}
  \end{center}
\end{table}

Once $E_{\rm vis}$ is given, the rate of the visible energy injection
during the BBN epoch is given by
\begin{equation}
  \label{eqiotanjection}
  \left[ \frac{d \rho_{\rm vis}}{dt} \right]_{\rm ann}
  = 
  \frac{1}{2}
  E_{\rm vis} n_{\rm DM}^2\langle \sigma v\rangle.
\end{equation}
With the given injection rate, the spectra of the energetic photon and
electron induced by the DM annihilation are obtained by solving a set
of Boltzmann equations, which include effects of various radiative
processes (photon-photon pair creation, inverse Compton scattering,
Thomson scattering and so on).  Then, we have incorporated the
photo-dissociation rates in network calculation of BBN and obtained
the abundances of the light elements.  In this calculation we have
used the most recent data for nuclear reaction
rates~\cite{Smith:1992yy,Angulo:1999zz,Cyburt:2001pp,Serpico:2004gx,
  Cyburt:2008up} and estimated the theoretical errors by Monte Carlo
method~\cite{Krauss:1990,Smith:1992yy}.

The details of the photo-dissociation effects have been discussed, for
example, in Ref.~\cite{Kawasaki:1994sc}.  The important points of the
photo-dissociation effects are summarized as follows:
\begin{itemize}
\item $T \gtrsim 10$~keV: The emitted high energy photons are quickly
  thermalized through photon-photon process and no significant
  photo-dissociation of the light elements occurs.
\item $1$~keV $\lesssim T \lesssim 10$~keV: The only D is destroyed by
  the high energy photons and its abundance decreases. The other light
  elements are not affected.
\item $T \lesssim 1$~keV: All light elements are destroyed by the high
  energy photons produced by DM annihilation. As a result, the
  abundance of $^4$He decreases while D, $^3$He and $^6$Li are
  non-thermally produced. In this case $^3$He to D ratio gives the 
  most stringent constraint.  
\end{itemize}

Here we make comments on annihilation onto hadronic particles.  Even
if DM annihilates mainly into charged leptons, we expect some emission
modes into hadrons via, for example, ${\rm DM}+{\rm DM} \rightarrow
\ell^{+}+\ell^{-} + Z^{*}$, where $\ell^\pm$ denotes charged lepton
hereafter.  As will be seen in next subsection, energetic hadrons,
especially nucleons, cause ``hadro-dissociation'' of the light
elements and give more serious effect on BBN than high energy
electrons and photons \cite{Jedamzik:2004er, Kawasaki:2004yh,
  Kawasaki:2004qu, Kawasaki:2008qe,Jedamzik:2006xz}.  However, the
expected branching ratio into hadrons is $\sim 10^{-3}$, which is too
small to give a significant constraint in the present problem.  (See
Fig.\ \ref{fig:BBNbb}.)

For annihilation into $\tau$'s, pions are produced from $\tau$
decays. Approximately 0.9 charged pions are emitted by the decay of
$\tau$ lepton in average.  As we will see in the next section, the
effect of pions might be important at around $T \sim 1$~MeV when the
$n/p$ is affected by $n + \pi^{+} \to p + \pi^0$ and $p + \pi^{-} \to
n + \pi^0$. The $p\leftrightarrow n$ inter-converting reactions
increase $n/p$, which results in larger value of the $^4$He abundance.
However, we have checked that the constraint from the overproduction
of $^4$He is much weaker than the constraint from the
photo-dissociation.

\subsubsection{Hadronic effects}

Next we discuss the effects of hadron emissions on the abundances of
the light elements.  As we will see, if the annihilation process has a
hadron-emission mode with a net branching ratio of $\gtrsim$ {\cal
  O}(10)$\%$, the most stringent constraint is mainly from the
hadronic modes. The exceptional case is that the emission time is very
late ($T \lesssim 0.3$ keV).

Once energetic colored particles are emitted by the annihilation, they
are hadronized and the energetic mesons and baryons are produced.  In
particular, once energetic nucleons are emitted after the
hadronization process, they may scatter off the background nuclei and
induce non-thermal production and destruction processes of light
elements.  The most significant processes are via the scattering off
the background ${\rm ^4He}$.  The effects of the non-thermal
production of the nuclei $A$ (= D, $^{3}$He, $^{4}$He, $^{6}$Li, and
$^{7}$Li) can be taken into account by including the following
production term into the Boltzmann equations:
\begin{eqnarray}
  \label{eq:n_hadronic}
  \left[ \frac{d n_A}{dt} \right]_{\rm ann}
  = \frac{1}{2} \xi_{A,{\rm ann}} n_{\rm DM}^2 
  \langle \sigma v\rangle.
\end{eqnarray}
Here, $\xi_{A,{\rm ann}}$ is the number of the produced nucleus $A$
per single annihilation process, which depends on the background
temperature and the energy spectrum of the emitted hadrons.  The
spectra of the produced hadrons depend on the annihilation mode and
the mass of the DM.  We note here that the mass-dependence is given by
$ \xi_{A,{\rm ann}}\propto m_{\rm DM}^{0.5\pm \delta}$ with $\delta
\lesssim 0.2$ \cite{Kawasaki:2004qu}.  Then, as we see in the
following, the constraint from the hadron emission approximately
scales as $\propto\langle\sigma v\rangle/m_{\rm DM}^{1.5}$.  Such
processes become effective when the cosmic temperature is lower than
$\sim 0.1\ {\rm MeV}$.

At higher cosmic temperature, another process, which is the
$p\leftrightarrow n$ conversion process induced by emitted pions, is
more effective.  If the cosmic temperature is sufficiently high
($T\gtrsim 0.1$\ MeV), all the emitted hadrons are stopped through
electromagnetic interaction with background electrons and photons
before they scatter off the background nuclei. In that case, the
stopped hadrons such as charged pions (as well as $n\bar{n}$ and
$p\bar{p}$) scatters off the background nuclei only through exothermic
reactions with their threshold cross sections. Then background neutron
and proton are inter-converted each other through the process like $n
+ \pi^{+} \to p + \pi^0$ and $p + \pi^{-} \to n + \pi^0$. Those
processes change the neutron to proton ratio even after the freezeout
time of neutron. In this case more $^{4}$He and D are produced through
the non-thermal interconversion
processes~\cite{Reno:1987qw,Kohri:2001jx}. Note that unstable hadrons
such as $\pi^{0}$ decay with short lifetimes ($\tau \ll 10^{-8}$ sec)
and disappear well before contributing to the inter-converting process.

We have included the relevant hadro-dissociation processes as well as
effects of $p\leftrightarrow n$ conversion into the BBN network
calculation and obtained the light element abundances.  The details
are discussed in \cite{Kawasaki:2004yh, Kawasaki:2004qu,
  Kawasaki:2008qe}, and here we just summarize the most important
features of the hadronic effects:
\begin{itemize}
\item $T \gtrsim 10$~MeV: The emitted hadrons cannot change the
  neutron to proton ratio and do not affect any 
  light element abundances at all.
\item 100~keV $\lesssim T \lesssim 10$~MeV: $^4$He and  D are produced
  through the $n \leftrightarrow p$  interconversion caused by stopped
  mesons ($\pi^{\pm}$) and/or nucleon-antinucleon pairs ($p\bar{p}$
  and $n\bar{n}$).
\item 10~keV $\lesssim T \lesssim$ 100~keV: $^4$He is destroyed by
  energetic nucleons. That produces a lot of energetic D, T and
  $^3$He. Among those daughter nuclei, D gives most stringent bounds on
  the amount of the primary nucleons emitted by the annihilation.
\item 0.3~keV~$\lesssim  T \lesssim 10$~keV: $^{6}$Li is non-thermally
  produced by the scattering between the background $^{4}$He and the
  energetic T. This gives the severest bound.  
\item $T \lesssim 0.3$~keV: The effect from the photo-dissociation can
  dominate and the $^{3}$He to D ratio gives the severest bound when
  $E_{\rm vis}/m_{\rm DM} \gtrsim 0.1$ even if the hadronic mode is
  at 100$\%$.
\end{itemize}

\subsection{Light element abundances: observations}

Next, we summarize observational constants on primordial abundances of
D, $^{3}$He $^{4}$He, $^{6}$Li and $^{7}$Li, which we adopt in our
study.  The errors are presented at 1 $\sigma$.  The subscript ``p''
and ``obs'' are for the primordial and observational values,
respectively.

The primordial value of D abundance is inferred from observation of
high redshift QSO absorption systems.  We adopt a following
observational constraints on the deuterium abundance as ``Low'' value:
\begin{eqnarray}
    \label{eq:Dobs}
    {\rm Low~ (n_{\rm D}/n_{\rm H})}_{\rm p} = 
    (2.82 \pm 0.26) \times 10^{-5},
\end{eqnarray}
which is the most-recently reported value derived by taking the
weighted mean of six observed QSO absorption
systems~\cite{O'Meara:2006mj}. This value well agrees with the baryon
to photon ratio suggested by the WMAP 5-year CMB anisotropy
observation~\cite{Komatsu:2008hk}. However, the six measurements have
a larger dispersion as expected. In addition, the deuterium is the
most fragile element.  So, in order to derive a conservative
constraint, we also adopt the highest value among the six measurements
as ``High'' D/H,
\begin{equation}
    \label{eq:Dobs_high}
    {\rm High~ (n_{\rm D}/n_{\rm H})}_{\rm p} = 
    (3.98^{+0.59}_{-0.67}) \times 10^{-5}.
\end{equation}

To constrain the primordial $^{3}$He abundance, we use an
observational $^{3}$He to D ratio as an upper bound, which is a
monotonically increasing function of the cosmic time.  (For details,
see Refs.~\cite{Sigl:1995kk,Kawasaki:2004qu}.)  In this study we adopt
the newest values of D and $^{3}$He abundances simultaneously observed
in protosolar clouds (PSCs), $(n_{\rm ^3He}/n_{\rm H})_{\rm PSC} =
(1.66 \pm 0.06) \times 10^{-5}$ and $(n_{\rm D}/n_{\rm H})_{\rm PSC} =
(2.00 \pm 0.35) \times 10^{-5}$~\cite{GG03}.  Then we get
\begin{eqnarray}
    \label{eq:He3D}
    (n_{\rm ^3He}/n_{\rm D})_{\rm p} < 0.83+0.27.
\end{eqnarray}

The mass fraction of $^{4}$He is determined by the measurement of
recombination lines from extragalactic HII regions. In the most recent
analysis~\cite{Izotov:2007ed}, two values are reported by using old
and new $^{4}$He-emissivity data, $Y_{\rm p}$ = $0.2472 \pm 0.0012$
and $Y_{\rm p}$ = $0.2516 \pm 0.0011$, respectively.  Notice that the
error presented in~\cite{Izotov:2007ed} does not include systematic
effects.  Thus, in this study, we add an error of $0.0040$
\cite{Fukugita:2006xy} to derive conservative constraint:
\begin{eqnarray}
    \label{eq:He4obs}
    Y_{\rm p} = 0.2516 \pm 0.0040.
\end{eqnarray}
(For the systematic uncertainty of the observed $^{4}$He abundance,
see also Ref.~\cite{Peimbert:2007vm}.)

For $^{7}$Li observed abundance in high temperature metal-poor
population II stars is considered as primordial. Here, we adopt the
most recent value of the $^{7}$Li to hydrogen ratio
$\log_{10}(^{7}$Li/H)$_{\rm obs}=-9.90\pm 0.09$ given in
Ref.~\cite{Bonifacio:2006au}.  This is close to the value given in
Ref.~\cite{RBOFN}: log$_{10}(^{7}$Li/H)$_{\rm obs}=-9.91\pm 0.10$. On
the other hand, a slightly larger value has been also reported in
Ref.~\cite{Melendez:2004ni}: log$_{10}(^{7}$Li/H)$_{\rm obs}=-9.63\pm
0.06$.  The problem is that the theoretical value of $^{7}$Li is much
larger than the observational one even if we adopt the higher value of
the observational abundance.  The situation becomes worse when we use
an updated reaction rate of
$^{4}$He($^{3}$He,$\gamma$)$^{7}$Be~\cite{Cyburt:2008up,Cyburt:2008kw}. The
observational values in Refs.~\cite{Bonifacio:2006au} and
\cite{Melendez:2004ni} are smaller than the standard BBN prediction by
approximately 0.35 dex and 0.25 dex, respectively.  As for $^{6}$Li
abundance, on the other hand, recent observation shows that the
theoretical value is much smaller than that of the observation,
($^{6}$Li/$^{7}$Li)$_{\rm obs}$ = 0.046 $\pm$
0.022~\cite{Asplund:2005yt}. These two discrepancies may be
collectively called ``lithium problem''. Concerning the inconsistency
between the face value of the primordial $^{7}$Li abundance and the
standard BBN prediction, various solutions have been discussed from
the viewpoints of both astrophysics
\cite{Pinsonneault:2001ub,Korn:2006tv,Li7prod_Astro} and cosmology
\cite{Li7prod_Cos,Jedamzik:2004er,Jedamzik:2006xz}.~\footnote{For
  references, see also \cite{astroLi6prod} for possible astrophysical
  scenarios to enhance $^{6}$Li abundance, and \cite{Cayrel:2007te}
  for its observational uncertainties.}  However, the main purpose of
this paper is to derive a conservative constraint, so we do not go
into the details of these models.  Instead, assuming some depletion,
we add an additional systematic error of +0.35 dex into the
observational face value $n_{^{7}{\rm Li}}/n_{\rm H}$ in our study:
\begin{eqnarray}
    \label{eq:Li7obs}
    \log_{10}(n_{^{7}{\rm Li}}/n_{\rm H})_{p}=
    -9.90 \pm 0.09 + 0.35.
\end{eqnarray}
Notice that this depletion factor $D_7=0.35$ is the systematic error a
little larger than that allowed from effect of $^7$Li depletion by
rotational mixing in stars~\cite{Pinsonneault:2001ub}.\footnote
{The depletion factor $D_7$ is defined as $D_7=\Delta \log_{10}(^7{\rm
    Li}/{\rm H})$. The depletion factor for $^6$Li is defined similarly.}

As for a $^{6}$Li constraint, we use $(n_{\rm ^6Li}/n_{\rm
  ^{7}Li})_{\rm obs}$ = 0.046 $\pm$ 0.022, which was newly-observed in
a very metal-poor star~\cite{Asplund:2005yt}. We also add a systematic
error of +0.106 \cite{Kanzaki:2006hm} to take into account depletion
effects in stars adopting the relation between $^7$Li and $^6$Li
depletion factors, $D_6 = 2.5 D_7$~\cite{Pinsonneault:1998nf,
Pinsonneault:2001ub},\footnote
{This relation is obtained for depletion by rotational mixing.  If the
depletion is caused by atomic diffusion~\cite{Korn:2006tv}, this
relation is no longer correct. Instead we should have used $D_6 \simeq
D_7$. However, in this case the upper bound on $^6$Li/$^7$Li becomes
more stringent than given by Eq.~(\ref{eq:Li6obs}). We prefer to use
the relation for the rotation mixing because we will get more
conservative constraint. }
which leads to $\Delta \log_{10}(n_{^{6}{\rm Li}}/n_{^{7}{\rm Li}}) =
0.525$ for $D_7=0.35$. Then we get a following upper bound:
\begin{eqnarray}
    \label{eq:Li6obs}
    \left( n_{\rm ^6Li}/n_{\rm ^{7}Li} \right)_p
     < 0.046 \pm 0.022 + 0.106.
\end{eqnarray}

\section{Constraints on Annihilation Cross Section} 
\label{sec:constraint}

Now, we show the constraints on the annihilation cross section of the
DM particle.  As shown in Eq.\ \eqref{fdot}, the amount of energetic
particles produced by the annihilation process is proportional to the
annihilation cross section, and hence the effects on the light element
abundances become more enhanced as the cross section becomes larger.
Since the light element abundances predicted from the standard BBN is
consistent with the observations, we obtain an upper bound on
$\langle\sigma v\rangle$ in order not to spoil the success of the BBN.

In this section, we calculate the abundances of light elements for
several different DM scenarios, taking account of non-thermal
production processes discussed in the previous section.  Then, we
compare the theoretical results with observational constraints to
derive the upper bound on $\langle\sigma v\rangle$.  As we will see
below, the constraints on the cross section depend on the dominant
annihilation process; if the DM dominantly annihilates into charged
leptons and/or photons, the photo-dissociation processes are the most
important effect, while the hadronic effects becomes more significant
when colored particles are produced by the annihilation.  In the
following, we consider several typical cases.

\subsection{Charged-leptonic modes}

The simplest way of explaining the PAMELA and ATIC anomalies is to
adopt the possibility that the DM dominantly annihilates into charged
lepton pair: ${\rm DM}+{\rm DM}\rightarrow \ell^++\ell^-$.  We start
with considering such a possibility.

\begin{figure}[t]
 \begin{center}
 \vspace{-1.0cm}
 \includegraphics[width=0.5\linewidth]{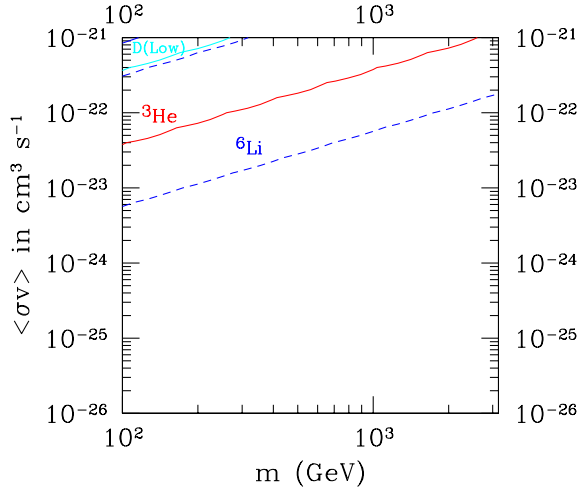} 
 \vspace{-1.0cm}
 \caption{\small BBN constraints on the annihilation cross section of
  DM particles when we take $D_{7}$ = 0.35. The red, cyan, blue,
  green and brown solid curves represent the upper bounds from
  observational constraints on $^3$He/D, D, $^6$Li, $^7$Li and
  $^4$He, respectively. The name of the element is also written by
  each line.  As for D, we plot both upper limits from High and Low
  values. Here we assume that the dark matter annihilates only into
  $e^+ e^-$, which means the fraction of the visible energy is
  $E_{\rm vis}/m=2$. The lines of $^4$He, $^6$Li and
    $^7$Li do not appear in this figure. For reference, the region
  sandwiched by two dashed lines is allowed by $^{6}$Li when we take
  $D_{7}$ = 0.}
  \label{fig:BBNrad}
 \end{center}
\end{figure}

We first show the constraints on $\langle\sigma v\rangle$ for the case
where ${\rm DM}+{\rm DM}\rightarrow e^++e^-$ is the dominant
annihilation process (Fig.~\ref{fig:BBNrad}).  Since $E_{\rm
  vis}\propto m_{\rm DM}$, the constraints depend on $\langle\sigma
v\rangle/m_{\rm DM}$.  The most stringent constraint comes from
$^3$He/D.

We can convert the constraints given in Fig.\ \ref{fig:BBNrad} to
other cases as far as the effects of hadronic decay modes are
negligible; since the effects of the photo-dissociation depend only on
$E_{\rm vis}$, the upper bounds from the photo-dissociation processes
scale as $E_{\rm vis}$.  In particular, if we use the most stringent
upper bound, which is from $^3$He/D, we obtain

\begin{equation}
  \label{eq:simgalep}
  \langle \sigma v\rangle < 3.5 \times 10^{-22}~ 
  {\rm cm}^3{\rm s}^{-1}\left(\frac{E_{\rm vis}}{2m_{\rm DM}}\right)^{-1}
  \left(\frac{m_{\rm DM}}{1\ {\rm TeV}}\right).
\end{equation}
For the case where ${\rm DM}+{\rm DM}\rightarrow\mu^++\mu^-$ or
$\tau^++\tau^-$ is the dominant annihilation process, the constraints
can be easily obtained from Fig.\ \ref{fig:BBNrad} by using Eq.\
\eqref{eq:simgalep}, because the photo-dissociation effect on BBN is
determined by the total visible energy injection by the annihilation
process.\footnote{
For $T \lesssim$ keV where the photo-dissociation of $^{4}$He is
effective, both muon and tau leptons decay well before they loose
their initial kinetic energies through scattering off the background
photons and electrons. That is because timescales of the Inverse
Compton (IC) and the Coulomb scatterings (CS) are given by $t_{\rm IC}
\sim 2.2\times 10^{-10}~{\rm s}~(\gamma_{\ell^\pm}/10^{3})^{-1}(T/{\rm
keV})^{-4}(m_{\ell^\pm}/m_{e})^{3}$ and $t_{\rm CS}\sim0.76\times
10^{5}~{\rm s}~(E_{\ell^\pm}/10^{2} {\rm GeV})(T/ {\rm keV})^{-3}$
with $\gamma_{\ell^\pm}$ being the Lorentz factor, $m_{\ell^\pm}$ the
mass, and $E_{\ell^\pm}$ the energy of charged lepton $\ell^\pm$ (=
$\mu$ and $\tau$), respectively.}

For the cases of ${\rm DM}+{\rm DM}\rightarrow\mu^++\mu^-$ and
$\tau^++\tau^-$, the upper bounds are given by $\langle \sigma v
\rangle < 1.0\times 10^{-21}\ {\rm cm^3 s^{-1}}\times (m_{\rm DM}/1\
{\rm TeV})^{-1}$ and $1.2\times 10^{-21}\ {\rm cm^3 s^{-1}} \times
(m_{\rm DM}/1\ {\rm TeV})^{-1}$, respectively.

\subsection{Hadronic modes}

Next, we consider effects of hadronic modes.  Even though we are
mainly interested in models which can explain the excess of cosmic-ray
$e^\pm$, it does not necessarily exclude models in which significant
amount of hadrons are emitted by the annihilation process.

One of the well-motivated cases is that the annihilation cross section
into the $W^{+}W^{-}$ pair is sizable.  This is the case when, for
example, the Wino in the supersymmetric model is the lightest SUSY
particle (LSP) and is the DM.  As we discuss in the following
subsection, the PAMELA and ATIC anomalies may be explained in such a
case.  Then, importantly, leptonic decay of the produced $W^\pm$ is
important for the production of cosmic-ray $e^\pm$, while the BBN
constraint is mainly due to the hadronic decay mode.

\begin{figure}[t]
 \begin{center}
 \vspace{-1.0cm}
 \includegraphics[width=0.5\linewidth]{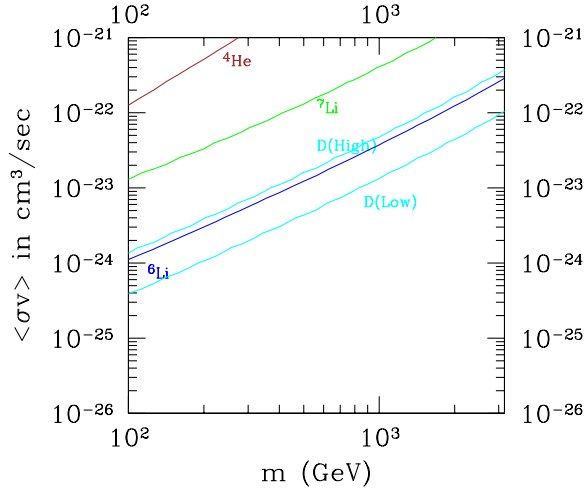} 
 \vspace{-1.0cm}
 \caption{\small Same as Fig.~\ref{fig:BBNrad} but for the annihilation
  into $W^{+}W^{-}$ pair. Then the fraction of the visible energy is
  $E_{\rm vis} / m = 0.94$. }
  \label{fig:BBNWW}
 \end{center}
\end{figure}

In Fig.~\ref{fig:BBNWW}, we show the bound on the annihilation cross
section for the case where the DM annihilates only into the
$W^{+}W^{-}$ pair.  For $m_{\rm DM}\lesssim 2\ {\rm TeV}$, as one can
see, the D abundance gives the severest constraint on the annihilation
cross section if we adopt the low value of the observational D; this
is due to the hadro-dissociation process of abundant ${\rm ^4He}$.
Then, in such a case, we can obtain an approximate constraint on the
annihilation cross section as a function of the mass: $\langle \sigma
v\rangle \lesssim 10^{-23}~ {\rm cm}^3{\rm s}^{-1}\left( N_{n}/0.8
\right)^{-1} \left(m_{\rm DM}/1\ {\rm TeV}\right)^{1.5}$, with $N_{n}$
being the number of emitted neutrons per single annihilation ($\sim$
0.8 in the case of $W^{+}W^{-}$ emission); here, we have used the
approximate relation discussed in the previous section: $\xi_{A,{\rm
ann}}\propto m_{\rm DM}^{0.5}$.  

\begin{figure}
 \begin{center}
    \includegraphics[width=0.5\linewidth]{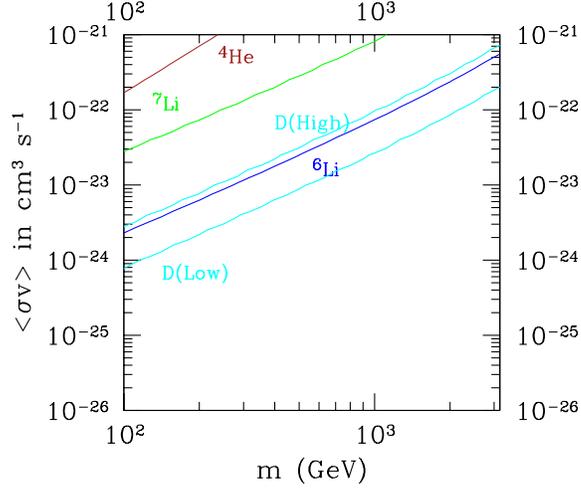}
    \includegraphics[width=0.5\linewidth]{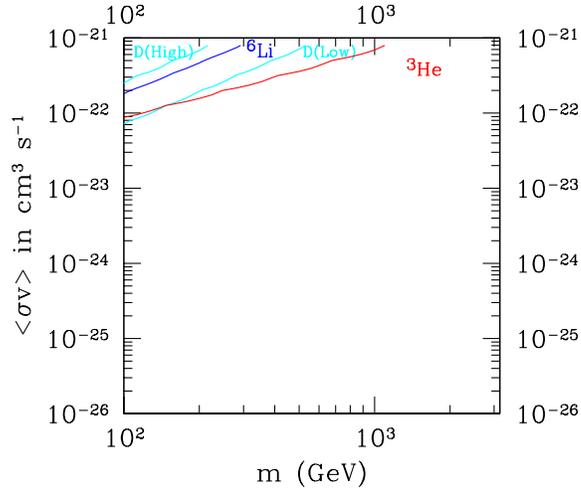}
    \caption{\small Same as Fig.~\ref{fig:BBNrad} but for the
    annihilation into $b\bar{b}$ pair (top panel). Then the fraction
    of the visible energy is $E_{\rm vis} / m =  1.04$. For
    comparison, we also plot the case of $b+\bar{b}$ emission at $1\%$
    in the bottom panel virtually by changing the hadronic branching
    ratio from 1 to 0.01 by hand with keeping the same value of
    $E_{\rm vis}$.}
  \label{fig:BBNbb}
 \end{center}
\end{figure}

So far, we have considered cases where energetic charged leptons (in
particular, $e^\pm$) are directly produced by the annihilation of the
DM or the decay of heavy particle (i.e., $W^\pm$).  Such cases are
important to explain the behavior of the $e^\pm$ fluxes observed by
the PAMELA and ATIC, because $e^\pm$ produced via the hadronization
process is less energetic.  Consequently, in the light of PAMELA and
ATIC, scenarios where the DM annihilates only into colored particles
(like a quark pair or a gluon pair) are less attractive.  However, in
general, the DM may annihilates dominantly into colored particles.
Thus, we also comment on constraints on the annihilation cross
section in such a case.  Here, as an example, we consider the case
where the DM annihilates dominantly into a $b\bar{b}$ pair; the
constraints is shown in the top panel of Fig.~\ref{fig:BBNbb}. The
fraction of the visible energy is $E_{\rm vis} / m_{\rm DM} =
1.04$. In this mode, the number of emitted neutrons is an increasing
function of $m_{\rm DM}$ (0.3--0.5 for $m_{\rm DM}$ = 100~GeV--1~TeV).
For comparison, we also study the case that $\langle\sigma
v\rangle_{{\rm DM}+{\rm DM}\rightarrow b+\bar{b}}/ \langle\sigma
v\rangle_{{\rm DM}+{\rm DM}\rightarrow {\rm all}}=0.01$, with keeping
the relation $E_{\rm vis}/m_{\rm DM}=1.04$; the constraints are shown
in the bottom panel of Fig.~\ref{fig:BBNbb}. Comparing the top panel
with the bottom one, we see that the constraints from the
photo-dissociation dominates if the hadronic mode is less than 1~$\%$.

\subsection{Implication to PAMELA/ATIC}

In previous subsections, we have seen that the quantity $\langle\sigma
v\rangle$ should be smaller than $\sim O(10^{-24}-10^{-21})\ {\rm
  cm^3s^{-1}}$ in order not to spoil the success of the BBN, if the
dominant annihilation process of the DM is into $e^+e^-$,
$\mu^+\mu^-$, $\tau^+\tau^-$, $W^+W^-$, $b\bar{b}$, and so on.  In
fact, the upper bound on $\langle\sigma v\rangle$ is of the similar order
of that required to explain the PAMELA and ATIC anomalies.

\begin{figure}
 \begin{center}
 \includegraphics[width=0.6\linewidth]{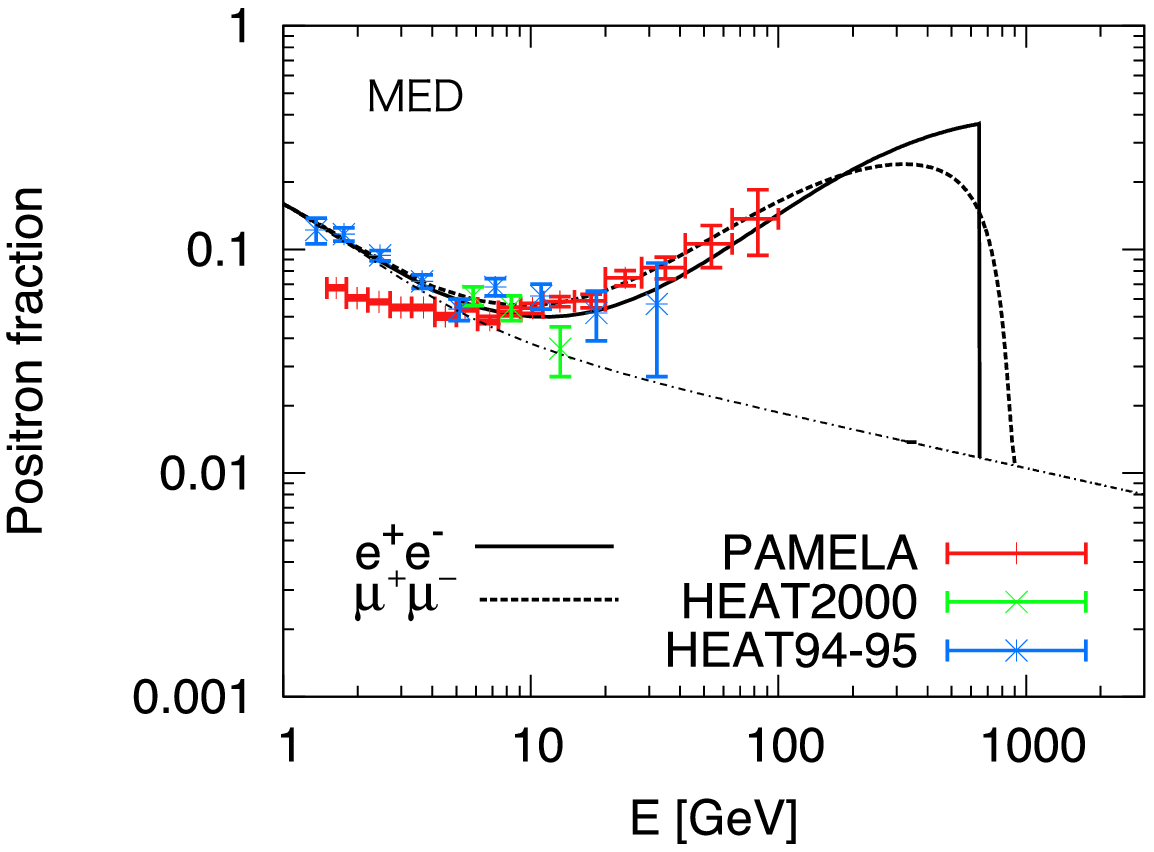}
 \includegraphics[width=0.6\linewidth]{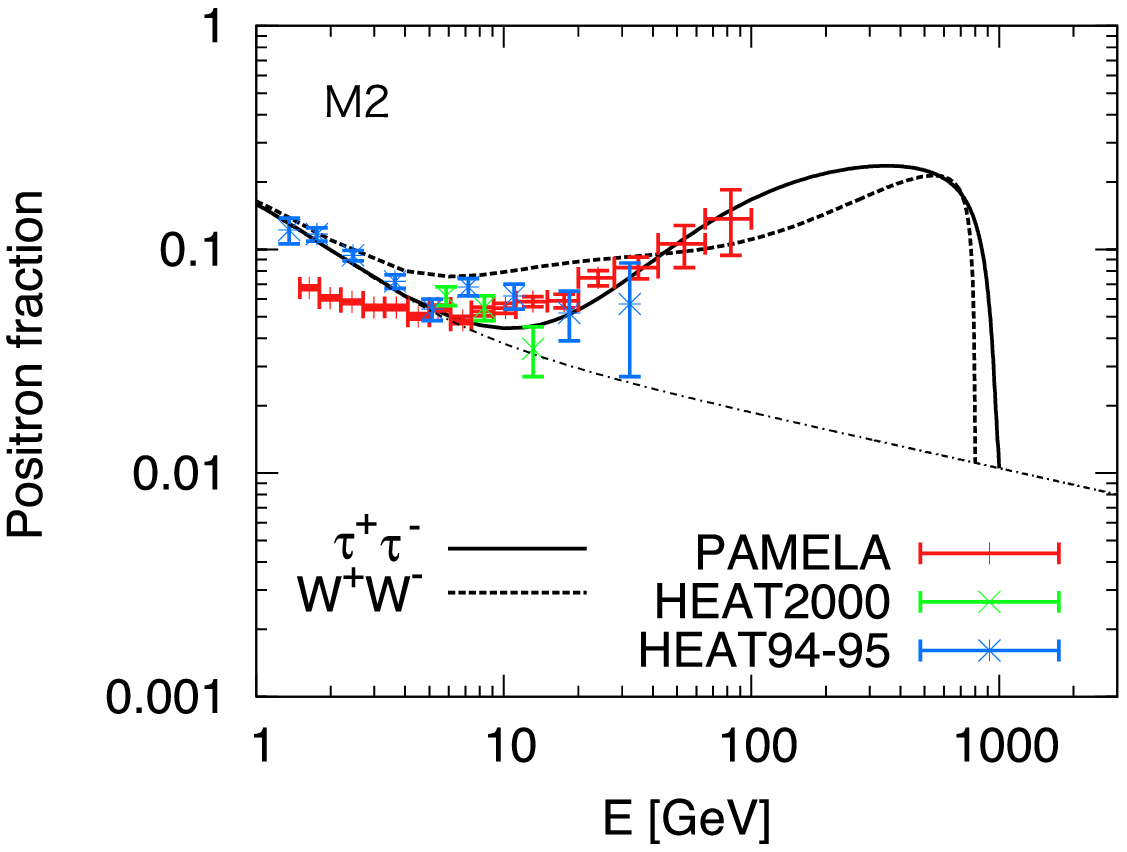}
 \caption{\small Positron fraction $R$ as a function of positron
   energy $E$.  (Top) We assume the DM annihilating into $e^+e^-$ with
   annihilation cross section $\langle \sigma v\rangle=5\times
   10^{-24}~{\rm cm^{3}s^{-1}}$ for $m_{\rm DM}=650~{\rm GeV}$, and
   into $\mu^+ \mu^-$ with $\langle \sigma v\rangle=15\times
   10^{-24}~{\rm cm^{3}s^{-1}}$ for $m_{\rm DM}=900~{\rm GeV}$ for the
   MED propagation model.  (Bottom) We assume DM annihilating into
   ${\tau}^+{\tau}^-$ with annihilation cross section $\langle \sigma
   v\rangle=4\times 10^{-23}~{\rm cm^{3}s^{-1}}$ for $m_{\rm
     DM}=1~{\rm TeV}$, and into $W^+W^-$ with $\langle \sigma
   v\rangle=3\times 10^{-23}~{\rm cm^{3}s^{-1}}$ for $m_{\rm
     DM}=800~{\rm GeV}$ for the M2 propagation model. Results of
   PAMELA and HEAT experiments are also shown.  }
  \label{fig:R}
 \end{center}
\end{figure}

\begin{figure}
 \begin{center}
 \includegraphics[width=0.6\linewidth]{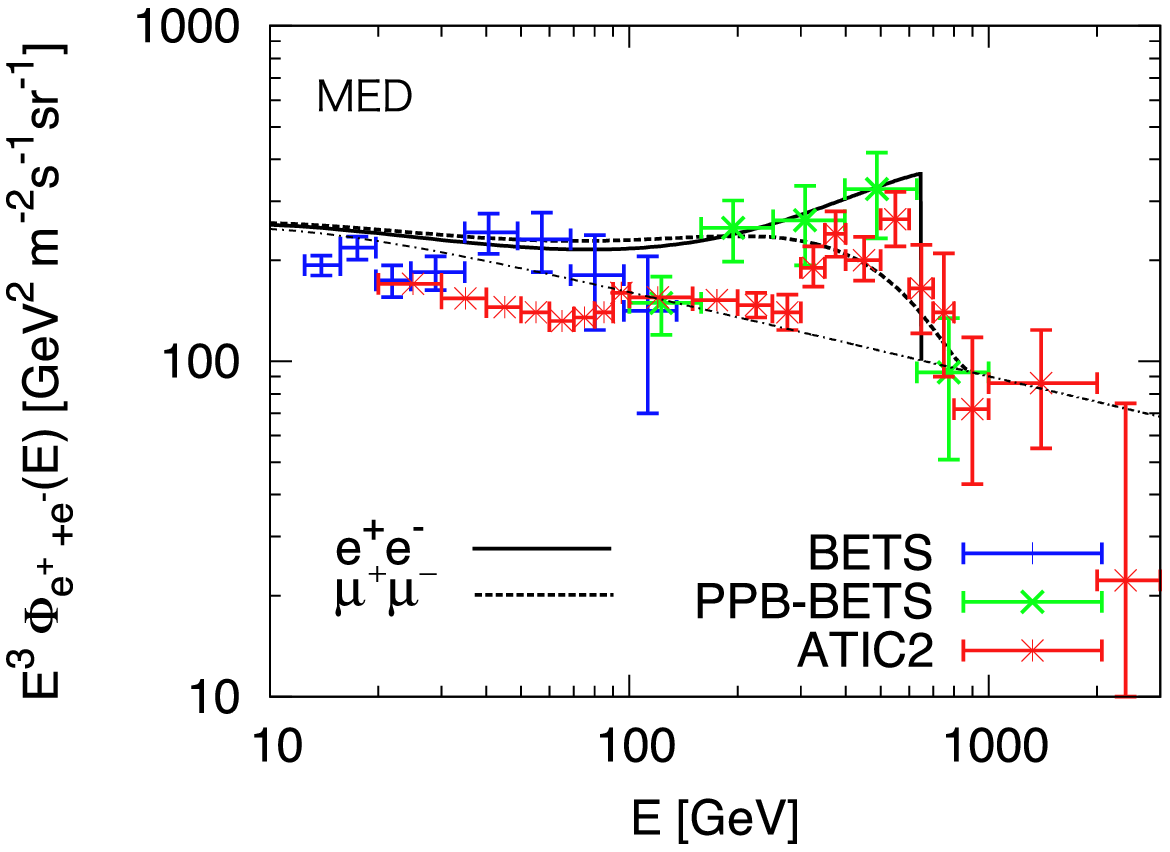}
 \includegraphics[width=0.6\linewidth]{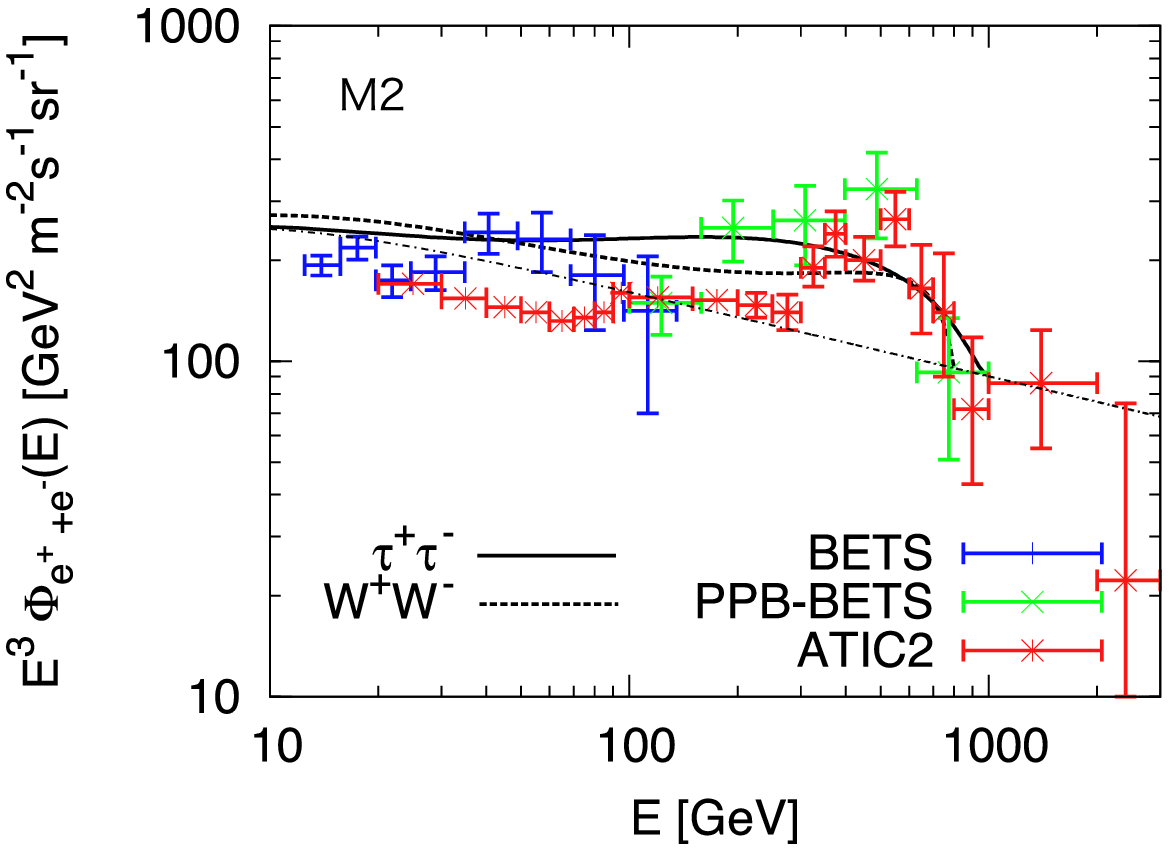}
 \caption{\small Total electron and positron flux (times $E^3$) as a
   function of their energy.  Model parameters are same as
   Fig.~\ref{fig:R}.  Results of BETS, PPB-BETS and ATIC are plotted.
 }
  \label{fig:eflux}
 \end{center}
\end{figure}

To see this, we plot the positron fraction and the total electron and
positron flux in Figs.\ \ref{fig:R} and \ref{fig:eflux}, respectively.
Here, we adopt the background fluxes of electron and positron of
Ref.~\cite{Moskalenko:1997gh}.  The annihilation cross section and the
DM particle mass (as well the propagation model) used in obtaining
Figs.\ \ref{fig:R} and \ref{fig:eflux} is summarized in Table\
\ref{table:model}.  In Fig.\ \ref{fig:R} (Fig.\ \ref{fig:eflux}), we
also show the data points of HEAT \cite{Barwick:1997ig} and PAMELA
\cite{Adriani:2008zr} (BETS \cite{Torii:2001aw}, PPB-BETS
\cite{Torii:2008xu} and ATIC \cite{:2008zz}).\footnote
{Recently the HESS collaboration also reported electron flux with
  $E\gtrsim 600~$GeV \cite{Collaboration:2008aa}, which is consistent
  with ATIC results.}
From the figures, we can see that DM models with the annihilation
cross section of $\langle \sigma v\rangle \sim 10^{-23}~{\rm
  cm^{3}s^{-1}}$ well explains the anomalies in the cosmic-ray $e^\pm$
fluxes.

To calculate the electron and positron fluxes from the DM
annihilation, we have adopted the conventional procedure
\cite{Hisano:2005ec}.  Approximating the shape of the diffusion zone
as a cylinder (with half-height $L$ and the length $2R=40\ {\rm
  kpc}$), we have derived the static solution to the following
diffusion equation \cite{Baltz:1998xv}:
\begin{equation}
   \frac{\partial}{\partial t}f(E, \vec x) = K(E)\nabla^2f(E, \vec x)
   +\frac{\partial}{\partial E} [b(E)f(E, \vec x)] + Q(E,\vec x), 
   \label{diffusion}
\end{equation}
where $f(E, \vec x)$ is the electron/positron number density with
energy $E$ at the position $\vec x$, which is related to the DM
contribution to the electron and positron fluxes as
\begin{equation}
   \Phi^{(\rm DM)}_{e^+}(E,\vec x_\odot)
   = \frac{c}{4\pi}f(E,\vec x_\odot),
\end{equation}
where $c$ is the speed of light and $\vec x_\odot$ denotes the
position of the solar system.  In Eq.\ (\ref{diffusion}), $K(E)$ is
the diffusion constant, and $Q(E,\vec x)$ is the source term from the
DM annihilation.  For the energy loss rate, we use $b(E)=1\times
10^{-16}(E/1~{\rm GeV})^2$.

The derived positron and electron fluxes are sensitive to the
diffusion constant and the half-height of the diffusion cylinder $L$.
The diffusion constant is parametrized as $K=K_0(E/1~{\rm
  GeV})^\delta$.  Hereafter we consider the following two propagation
models, called the MED model: $(K_0, \delta, L)=(0.0112~{\rm
  kpc^2/Myr}, 0.70, 4~{\rm kpc})$ and M2 model: $(K_0, \delta,
L)=(0.00595~{\rm kpc^2/Myr}, 0.55, 1~{\rm kpc})$, both of which are
consistent with observed boron-to-carbon ratio in the cosmic-ray flux
\cite{Delahaye:2007fr}.

The source term in Eq.~(\ref{diffusion}) is given by
\begin{equation}
  Q(E,\vec x) = \frac{1}{2} B_F n_{\rm DM, now}^2(\vec x) 
  \langle \sigma v \rangle
  \left[ \frac{dN_{e^\pm}}{dE} \right]_{\rm ann},
  \label{sourceterm}
\end{equation}
where $n_{\rm DM, now}(\vec x)$ is the present DM distribution (in
particular, in the halo).  Dependence of the electron and positron
fluxes on the DM density profile is very minor.  In our calculation,
we use the isothermal density profile as $\rho_{\rm DM,now} (r)=
m_{\rm DM}n_{\rm DM,now}(r)=0.43(2.8^2+8.5^2)/(2.8^2+r_{\rm
  kpc}^2)~{\rm GeV~cm^{-3}}$, where $r_{\rm kpc}$ is the distance from
the Galactic center in units of kpc.  Furthermore, according to
$N$-body simulations, the dark matter may not be distributed smoothly
in our Galaxy and there may be clumpy structures somewhere in the
Galactic halo.  If this is the case, the positron flux may be enhanced
\cite{Silk:1992bh}.  In Eq.\ \eqref{sourceterm} such an effect is
considered by the boost factor $B_F$.

\begin{table}[t]
  \begin{center}
    \begin{tabular}{llll}
      \hline \hline
      Final State & $m_{\rm DM}\ ({\rm GeV})$
      & $B_F\langle\sigma v\rangle\ ({\rm cm^3s^{-1}})$
      & Propagation Model
      \\
      \hline
      $e^+ e^-$        & 650  & $0.5\times 10^{-23}$ & MED \\
      $\mu^+ \mu^- $   & 900  & $1.5\times 10^{-23}$ & MED \\
      $\tau^+ \tau^- $ & 1000 & $4.0\times 10^{-23}$ & M2 \\
      $W^+ W^-$        & 800  & $3.0\times 10^{-23}$ & M2 \\
      \hline \hline
    \end{tabular}
    \caption{\small DM particle mass, the product $B_F\langle\sigma
    v\rangle$, and the propagation model used in Figs.\ \ref{fig:R}
    and \ref{fig:eflux}.}
    \label{table:model}
  \end{center}
\end{table}


It is found from the figures that the best-fit values of the DM
particle mass and the annihilation cross section to the PAMELA and
ATIC observations depend on the annihilation modes.  For example, if
the DM dominantly annihilates as ${\rm DM}+{\rm DM}\rightarrow
e^++e^-$, $m_{\rm DM}\simeq 650\ {\rm GeV}$ and $B_F\langle\sigma
v\rangle\simeq 0.5\times 10^{-23}\ {\rm cm^3s^{-1}}$ is preferred to
explain the anomalies.\footnote
{ Model parameters used here and hereafter are just for the
  representative purpose and the fit to the observational data is
  still reasonable even if the parameters are varied up to a few ten's
  of percent.  However, the following conclusions are not affected
  with such changes of parameters.}
As we have seen in the previous subsection, when $m_{\rm DM}\simeq
650\ {\rm GeV}$, the BBN constraint requires $\langle\sigma
v\rangle\leq 2.3\times 10^{-22}\ {\rm cm^3s^{-1}}$, and hence the
PAMELA and ATIC anomalies can be explained even if $B_F=1$.
For the case where the DM
annihilates as ${\rm DM}+{\rm DM}\rightarrow W^++W^-$, it is
remarkable that we need a sizable boost factor (for example, $\gtrsim
4$), because the observations require $\langle \sigma v\rangle=3\times
10^{-23}~{\rm cm^{3}s^{-1}}$ for $m_{\rm DM}=800~{\rm GeV}$ for the M2
propagation model, as shown in Figs.~\ref{fig:R} and \ref{fig:eflux}.

\section{Conclusions and Discussion} \label{sec:conc}

In this paper we have investigated the effect of the DM annihilation
on BBN.  Recent observations of the cosmic-ray positron/electron
excess by the PAMELA/ATIC experiments can be interpreted as the
contribution from the DM annihilation.  However, this requires very
large annihilation cross section as $\langle \sigma v\rangle \sim
10^{-23}~{\rm cm^3s^{-1}}$ unless the boost factor is not introduced,
and such large annihilation rate can have significant effects on BBN.

We found that leptonically annihilating DM models are constrained from
the upper bound on $^3$He/D.  This limit is  compatible with
the DM annihilation models for explaining the PAMELA and ATIC
anomalies for the case of annihilation into $e^+e^-$, $\mu^+\mu^-$ and $\tau^+\tau^-$. 
On the other hand, the DM annihilation models with
significant amount of hadrons are constrained from the observations of
D or $^6$Li abundances.  This gives a stringent constraint in some
case; the DM models annihilating into $W$-bosons for explaining both
the PAMELA and ATIC anomalies are excluded unless the boost factor
larger than unity is introduced.

In this paper we assumed that the DM annihilation cross
section is independent of the cosmic time. 
However, our results also have impacts on
scenarios with the time-dependent cross section.  When the DM
annihilates through a pole just below the threshold, the
annihilation cross section is enhanced for lower relative velocity
of the annihilating DM particles \cite{Ibe:2008ye}.  This
enhancement is automatically implemented in the Sommerfeld mechanism
\cite{Hisano:2003ec,Hisano:2006nn}.  Then, the large boost
factor may not be required to realize both the thermal freezeout
scenario of the DM and the enhanced $e^\pm$ fluxes consistent with
the PAMELA/ATIC.  However, in such scenarios,
it is notable that the annihilation cross section
during the BBN epoch may be significantly enhanced than that in the
present Galaxy.  Since our results indicate that the DM
annihilation cross section in the BBN epoch cannot be much larger
than that in the Galaxy now, those models are excluded if the
enhancement of the cross section during the BBN epoch is sizable. 

Some comments are in order. The DM annihilation in the Galactic center
yields significant amount of gamma-rays through the cascade decay of
the annihilation products and/or bremsstrahlung processes, which must
be compared with the HESS observation \cite{Aharonian:2004wa}.
However, the gamma-ray flux significantly depends on the DM density
profile, and if a moderate profile such as an isothermal one 
with rather large core radius \cite{Salucci:2000}
is chosen, the constraint is loosen \cite{Bertone:2008xr}.

DM annihilation cross section can also be constrained from the
observations of anti-protons \cite{Donato:2008jk,Grajek:2008pg} in the
case of hadronic annihilation mode, and synchrotron radiation
\cite{Bertone:2008xr,Zhang:2008tb} produced by the DM annihilation
inside the Galaxy.  However, these constraints more or less suffer
from astrophysical uncertainties, such as the density profile of the
DM halo, the size of the diffusion zone and the distribution of the
magnetic field. They lead to orders of magnitude uncertainties on the
resulting constraints \cite{Donato:2003xg,Borriello:2008gy}.  Taking
into account these uncertainties, DM annihilation models are
consistent with observations \cite{Grajek:2008pg,Bertone:2008xr}.

Neutrinos from the DM annihilation at the Galactic center
\cite{Hisano:2008ah} (and possibly from the DM trapped inside the
Earth \cite{Delaunay:2008pc}) may also be useful as a tool for
cross-checking signatures of DM annihilation and Super-Kamiokande
\cite{Desai:2004pq} gives constraints on some DM annihilation models.
However, it still depends on the DM density profile, though the
dependence is rather weak.

The constraints on the DM annihilation cross section from BBN
presented in this paper is robust, since they do not suffer from large
astrophysical uncertainties, compared with those DM signatures in the
cosmic rays.

\section*{Acknowledgment}


K.N. would like to thank the Japan Society for the Promotion of
Science for financial support.  This work is supported by Grant-in-Aid
for Scientific research from the Ministry of Education, Science,
Sports, and Culture (MEXT), Japan, No.\ 20244037, No.\ 2054252 (J.H.),
No.\ 14102004 (M.K.)  and No.\ 19540255 (T.M.), and also by World
Premier International Research Center Initiative, MEXT, Japan.
K.K. is supported in part by STFC grant, PP/D000394/1, EU grant
MRTN-CT-2006-035863, the European Union through the Marie Curie
Research and Training Network ``UniverseNet.''




\end{document}